\documentclass[12pt,onecolumn,amsmath,amssymb]{revtex4}
\usepackage{graphicx}
\begin{document}

\title{Equilibrium Configurations of the Charged Surface
of a Conducting Liquid at a Finite Interelectrode Distance}
\author{Nikolay M. Zubarev and Olga V. Zubareva}
\email{nick@ami.uran.ru}

\affiliation{Institute of Electrophysics, Ural Branch, Russian
Academy of Sciences,\\ 106 Amundsen Street, 620016 Ekaterinburg, Russia}

\begin{abstract}
The problem of determining equilibrium configurations of the free surface
of a conducting liquid is considered with allowance for a finite
interelectrode distance. The analogy is established between this
electrostatic problem and that of finding the profile of a progressive
capillary wave on the free surface of a liquid layer of a finite depth,
which was solved by Kinnersley. This analogy allowed exact solutions to
be obtained for the geometry of liquid electrodes, which expand the
existing notions about the possible stationary states of the system.
\end{abstract}

\maketitle

As is known, flat surface of a conducting liquid
exposed to a sufficiently strong electric field becomes
unstable \cite{1}. This instability is induced by the Coulomb
forces, whereas capillary forces play a stabilizing role.
In order to understand the main laws governing the
behavior of such systems, it is necessary to establish
both the conditions under which the mutual compensation
of these forces is possible and the conditions where
this is basically impossible. This, in turn, makes necessary
an analysis of the possible equilibrium configurations
of the charged surface of liquid electrodes.

Previously \cite{2, 3}, exact solutions were obtained for
the equilibrium configuration of the surf ace of a conducting
liquid in a homogeneous electric field. In application
to analysis of the possible configurations of liquid
electrodes, this situation corresponds to the formal
limit of infinite interelectrode distances. The analysis
in \cite{2, 3} was based on the established analogy with the
problem of description of progressive capillary waves
on the free surface of a deep ideal liquid, which was
solved by Crapper in 1957. The form of equations for
the two-dimensional potential of the electric field and
the hydrodynamic current function in these problems
coincide to within the notation.

This paper will demonstrate that, by expanding the
aforementioned analogy to the case of a finite distance
between electrodes (and for capillary waves, to the case
of a liquid layer of a finite depth), it is possible to construct
exact solutions of the classical electrostatic problem
for the case of finite geometry. The corresponding
solutions for progressive capillary waves were found in
l976 by Kinnersley \cite{4}; in 1999, these solutions were
obtained using a more rational method by Crowdy \cite{5}.

Now we will write a set of equations determining
the equilibrium shape of the free surface of a conducting
liquid for a given interelectrode distance $d$ and a
potential difference $U$. Let the vector of the electric
field strength to be directed along the $y$ axis of a Cartesian
coordinate system. In the unperturbed state, the
liquid boundary is a flat horizontal surface $y\!=\!-d$, and
the position of the upper (flat solid) electrode corresponds
to $y\!=\!0$. Restricting the consideration to the
case of planar symmetry, we can describe the shape of
a perturbed surface of the liquid electrode by the function $\eta(x)$.
For incompressible liquids, the periodic solutions
must obey the condition $d\!=\!-\!\lambda^{-1}\!\int_0^\lambda \eta(x)\,dx$,
where $\lambda$ is the period. The electric field potential $\Phi$
is
described by the Laplace equation
$$
\Phi_{xx}+\Phi_{yy}=0,
$$
which has to be solved together with the boundary conditions
$$
\Phi=0, \qquad y=0,
$$
$$
\Phi=U, \qquad y=\eta(x).
$$
The equilibrium relief of the liquid boundary is determined
by the condition of balance of the forces acting
upon this surface:
$$
\frac{{\Phi_x}^2+{\Phi_y}^2}{8\pi}
+\frac{\alpha \eta_{xx}}{(1+{\eta_x}^2)^{3/2}}=p, \qquad y=\eta(x),
$$
where the first term in the left-hand part describes the
electrostatic pressure and the second term, the surface
pressure ($\alpha$ is the coefficient of surface tension and
$p$ has a meaning of the difference between the external
and internal pressures).

Comparing the above equations to those \cite{4} determining
the shape of a capillary wave in the system of
coordinates moving at a phase velocity $c$ with this wave,
we establish that these equations coincide to within the
substitution
$$
p\to\rho c^2/2,\quad
\Phi\to\sqrt{4\pi\rho}\,\Psi,\quad
\eta\to-\eta, \quad
y\to-y.
$$
Here $\Psi$ is a current function harmonically conjugated
with the velocity potential and $\rho$  is the liquid density.
Using the known symmetric solutions of these equations
(which correspond to case Ib in the notation of
Kinnersley \cite{4}), we obtain the following explicit parametric
expressions describing the electric field strength
distribution in the interelectrode gap:
\begin{equation}
x\!=\!\frac{\alpha}{2p{k'}^2}\!\left[2E(\psi,k)\!-\!{k'}^2\psi\!-\!
2k^2\mbox{sn}(\psi,k)\,\mbox{cd}(\psi,k)
\!+\!\frac{2k{k'}^2\mbox{sd}(\psi,k)\,
\mbox{nd}(\psi,k)}{\mbox{dn}(\varphi,k')-
k\,\mbox{cd}(\psi,k)} \right],
\end{equation}
\begin{equation}
y=\frac{\alpha}{2p{k'}^2}\left[(1+k^2)\varphi-2E(\varphi,k')
+\frac{2{k'}^2\mbox{sn}(\varphi,k')\,\mbox{cn}(\varphi,k')}
{\mbox{dn}(\varphi,k')-k\,\mbox{cd}(\psi,k)} \right].
\end{equation}
Here $\mbox{sn}$, $\mbox{cn}$, $\mbox{dn}$, $\mbox{sd}$, $\mbox{cd}$,
$\mbox{nd}$ are the Jacobi elliptic functions; $E$ is the incomplete
elliptic integral of second kind; $k$ is the modulus of the elliptic
integral; $k'\!=\!\sqrt{1-k^2}$ is the complementary modulus;
$\varphi\!=\!\sqrt{p/(2\pi\alpha^2)}\,\Phi$ is the dimensionless
electric field potential, and $\psi$ is the corresponding harmonically
conjugated function. On the liquid surface, potential $\varphi$
acquires the value $u\!=\!\sqrt{p/(2\pi\alpha^2)}\,U$,
so that the condition $\varphi\!=\!u$ determines the unknown
equilibrium surface in the parametric form ($\psi$ plays the role
of a parameter).

\begin{figure}
\includegraphics{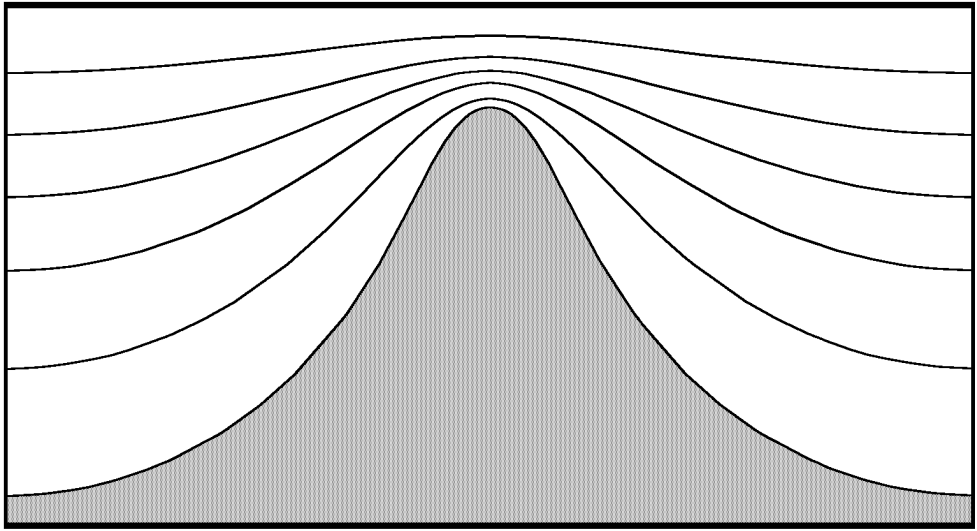}
\caption{\label{fig:figure}
One period of a stationary profile of the free surface of a
conducting liquid for $k = 0.2$ and $u = 1.8$. The curves show
equipotential surfaces $\varphi= 0.3, 0.6, 0.9, 1.2, 1.5$.
}
\end{figure}

It should be noted that the equilibrium configurations
of a charged liquid were previously studied only
in a weakly nonlinear limit, whereby the wavelength
was much greater than the amplitude of the surface
deformation (see, e.g. \cite{6,7,8} and references therein).
Using Eqs. (1) and (2), it is possible to analyze (without
allowance for the gravity field) the possible substantially
nonlinear surface configurations for which the
wavelengths and amplitudes are comparable (see figure~\ref{fig:figure}).
Important distinction of such an analysis from that
of Kinnersley is the basically different parametrization
of solutions. In the hydrodynamic problem, the main
control parameter was the phase velocity; in our case,
an analog of this velocity has mo physical meaning. In
the electrostatic problem, the control parameters are the
potential difference $U$ and the interelectrode distance $d$.
The latter quantity does not explicitly enter into the
expressions for solutions of Eqs. (1) and (2) and can be
calculated using the formula
\begin{equation}
d=-\lambda^{-1}\!\!\!\int\limits_{0}^{4K(k)}\!\!
\left.\left(x_\psi\,y\right)\right|_{\varphi=u}\,d\psi.
\end{equation}
Convenient parameters characterizing the solutions are
offered by the wavelength
\begin{equation}
\lambda=\frac{2\alpha}{p{k'}^2}
\left[2E(k)-{k'}^2K(k)\right],
\end{equation}
(where $K(k)$ and $E(k)$ are the elliptic integral of the first
and second kind, respectively) and the amplitude of the
surface perturbation
\begin{equation}
A=\frac{1}{2}\!\left.(y_{\max}-y_{\min})\right|_{\varphi=u}=
\frac{\alpha k}{p{k'}^2}\,\mbox{sc}(u,k').
\end{equation}
Excluding the moduli $k$, $k'$ and the pressure difference $p$
from relations (3)-(5), we obtain the dependence of the
stationary wave amplitude $A$ on the wavelength $\lambda$ and the
system parameters (the potential difference $U$ and the
interelectrode distance $d$). Analysis of this dependence,
which is beyond the framework of this short communication,
will allow us to study qualitatively the obtained
solutions with respect to their stability and to formulate
criteria for the growth of perturbations on the charged
surface of liquid electrodes.

The work was supported in part
by the Presidential grand (project no. MK-2149029942),
the Foundation for Support of Russian Science, the
"Dynasty" Foundation of Noncommercial Programs,
and the International Center of Basic Physics (Moscow).

\end{document}